%% file: mt_07.tex
\documentclass{aa}
\usepackage{apj2aa}
\usepackage{astron} 
\usepackage{float,epsfig,psfig}
\usepackage{pstricks}

\input {mydefs.tex}

\newfont{\sfsl}{cmssqi8 scaled 1200}
\newcommand{\gcs}{{\sfsl HIFLUGCS}}
\newcommand{\rhoc}{\rho_{\rm crit}}

\begin{document}
\thesaurus{11.03.1, 12.03.3, 12.04.1, 12.12.1
}

\title{{Details of the mass--temperature relation for clusters of galaxies}} 
\author{Alexis Finoguenov, Thomas H. Reiprich and Hans B\"ohringer}
\institute{Max-Planck-Institut f\"ur extraterrestrische Physik,
             Giessenbachstra\ss e, 85748 Garching, Germany}
\date{Received October 9, 2000; accepted: December 20, 2000}
\maketitle
\markboth{Finoguenov, Reiprich \& B\"ohringer: $M-T$ Relation}
{Finoguenov, Reiprich \& B\"ohringer: $M-T$ Relation}

\begin{abstract}
  
We present results on the total mass and temperature determination using two
samples of clusters of galaxies. One sample is constructed with emphasis on
the completeness of the sample, while the advantage of the other is the use
of the temperature profiles, derived with ASCA. We obtain remarkably similar
fits to the $M-T$ relation for both samples, with the normalization and the
slope significantly different from both prediction of self-similar collapse
and hydrodynamical simulations. We discuss the origin of these discrepancies
and also combine the X-ray mass with velocity dispersion measurements to
provide a comparison with high-resolution dark matter simulations. Finally,
we discuss the importance of a cluster formation epoch in the observed $M-T$
relation.
  
\end{abstract}


\section{Introduction}

There is increasing interest in using galaxy clusters characterized by X-ray
observations as probes for cosmic structure and its evolution. In one of the
applications, Press-Schechter models (Press \& Schechter 1974) are used to
constrain the shape and amplitude of the primordial density fluctuation
spectrum at present day scales of $5 - 10 h^{-1}$ Mpc (\eg Henry \& Arnaud
1991, Oukbir \& Blanchard 1992, Henry \etal 1992, Eke \etal 1996, 1998,
Borgani \etal 1999). Another important measurement is that of the spatial
correlations of X-ray clusters or, alternatively, the measurement of the
power spectrum of the density fluctuations in the cluster density
distribution (\eg Nichol \etal 1992, Romer \etal 1994, Retzlaff \etal 1998,
Miller \& Batuski 2000, Schuecker \etal 2000, Collins \etal 2000). In both
cases, the masses of the clusters have to be known for a large test sample
in order to apply it to the structure analysis. The mass determination is
still only feasible for a limited number of well-observed galaxy clusters,
however. Therefore, in the above approaches the empirically-found
correlation of the mass with the more easily observable parameters, X-ray
luminosity or intracluster gas temperature, are used to make the connection
between theoretical modeling and observations. The mass-temperature relation
plays a crucial role in this analysis, since it was found to be particularly
tight. Since the time of the first X-ray temperature measurements it was
observed that the X-ray temperature is closely connected to the velocity
dispersion of galaxies in clusters, indicating that both cluster components
trace the cluster potential in a similar way (\eg Mushotzky \etal 1978,
Mushotzky 1984).

The validity of this approach was investigated in numerical calculations
based on $N$-body dynamic and hydrodynamic simulations making predictions on
the mass-temperature relation and its dispersion, as measured in X-ray
studies (Evrard, Metzler, Navarro 1996, Evrard 1997). A close correlation
with a dependence of $M \propto T^{3/2}$ was predicted. It was found in this
work and also by \eg Schindler (1996) that the X-ray mass determination
should be reliable (with an uncertainty in the range $14-29$\%). Evrard
\etal (1996 and Evrard 1997) argue, however, that the predicted
mass-temperature relation has such a small dispersion that a mass estimate
based on the temperature measurement only is more accurate than that
determined on the basis of the additional knowledge of the gas density
profile as obtained from $\beta$-model surface brightness profile fits.
These predictions from simulations have been tested recently by means of
observational results by several authors (\eg B\"ohringer 1995, Hjorth \etal
1998, Horner \etal 1999, Nevalainen \etal 2000). Disagreements have been
found concerning the predicted slope (see \eg Horner \etal 1999, Nevalainen
\etal 2000) and the normalization of the mass-temperature relation.

At the time of writing, this discrepancy has not been solved, and various
aspects of the data analysis and simulations have been discussed: the
correct measurement of temperature gradients in the intracluster gas (\eg
Markevitch \etal 1998; Irwin, Bregman, Evrard 1999, White 2000), influence
of cooling flows (Allen 1998), non-thermal pressure support (\eg Schindler
1996; Fukazawa \etal 2000), influence of the heating of the intracluster
medium by supernovae (\eg Balogh, Babul, Patton 1999; Loewenstein 2000;
Finoguenov, Arnaud \& David 2000, hereafter FAD), density and velocity
bias in the galactic component of the cluster (Col\'{\i}n \etal 1999) and
effects of numerical resolution in the simulations (Nevalainen \etal 2000).

In this paper we reinvestigate the observational mass-temperature relation
using an extended sample of clusters with measured temperature profiles
covering a wide range of systems with luminosity-averaged temperatures from
below 1 up to 10 keV. In particular, we include 22 low temperature systems,
thus improving on sampling the low-mass part of the $M-T$ relation, compared
to previous studies (\eg\ Horner \etal 1999; Nevalainen \etal 2000).  To
study in detail the parameters that influence the normalization and the
slope of the $M-T$ relation, we rederive the $M-T$ relation assuming an
isothermal temperature distribution and also compare with the $M-T$ relation
for \gcs, a statistically complete sample of clusters, described in more
detail below, where the isothermality assumption was the only choice. This
sample is included in our study primarily to demonstrate that the result of
the $M-T$ relation is not affected by any selection bias. In addition, we
discuss implications from the results of the $M-T$ relation study on the
epoch (redshift) of cluster formation.

The paper is organized as follows. In section \ref{sec:data} we describe the
sample compilation, in \ref{sec:poly} we compare the results obtained for
the two different samples, in \ref{sec:opt} we combine the X-ray mass with
velocity dispersion measurements to provide a comparison with
high-resolution dark matter simulations and in \ref{sec:zf} we discuss a
correction for the redshift of cluster formation. Unless noted otherwise, we
assume $\Omega_{\rm M}=1$, $\Lambda=0$ and \h0, $\rho_{\rm crit,
o}=4.6975\times10^{-30}$ g cm$^{-3}$ throughout the paper.

\section{Data Compilation and Results}\label{sec:data}

In this paper we compare the $M-T$ relation of two cluster samples. The
first is a statistically complete sample of the X-ray brightest clusters,
the HIghest X-ray FLUx Galaxy Cluster Sample (\gcs ), for which the
selection criteria and thus possible bias effects are well defined. For this
sample, masses have been determined on the assumption of isothermality of
the X-ray emitting plasma. The second sample comprises those clusters for
which we have temperature profiles from ASCA observations and for which a
refined mass determination is possible. For the analysis of both samples, we
use the same definition of the ``total mass'' as the mass enclosed by the
radius, $r_{500}$, inside which the mean cluster mass density is 500 times
higher than the critical density of the Universe. Accounting for the
observed redshift means that the critical density, used to calculate the
overdensity, is scaled for every cluster according to its redshift. Later,
in fitting the $M-T$ relation, we correct the temperature of the cluster by
dividing by $(1+z)$. These corrections mean that clusters form at different
epochs at similar overdensity, so \( M 3 / (4\pi R^3) = 180 \rho_{\rm crit,
o} (1+z)^3 \). Therefore, for a fixed mass $R \propto (1+z)^{-1}$, and
$T\propto M/R \propto (1+z)$. This is strictly correct only for $\Omega_{\rm
M}=1$. However, correction for the observed redshift assuming lower values
of $\Omega_{\rm M}$ has even smaller effect on the derived parameters of the
$M-T$ relation, so our correction, assuming $\Omega_{\rm M}=1$, can be
considered as an extreme case. As we will show below, even in this case,
there is no difference in the results obtained for both cluster samples
considered here and thus no difference in the derived parameters of the
$M-T$ relation is expected for any value of $\Omega_{\rm M}$.

\subsection{\gcs}

Candidates for \gcs\ have been selected from recent cluster catalogs based
on the ROSAT All-Sky Survey. They have been reanalyzed homogeneously to
construct a complete X-ray flux-limited sample of the brightest clusters in
the sky, comprising 63 clusters. Details of the sample construction are
described in Reiprich \etal (2001). The cluster masses have been calculated
by assuming hydrostatic equilibrium and isothermality of the intracluster
gas and determining the gas density profile with the $\beta$ model
(Cavaliere and Fusco-Femiano 1976, Gorenstein \etal 1978, Jones and Forman
1984). Overall cluster temperatures have been compiled from the literature,
giving preference to temperatures measured with the ASCA satellite. The
cluster masses, $M_{500}$ and the mass errors, determined by adding the
temperature error and the errors of the fit parameter values ($\beta$ and
core radius), are tabulated in Reiprich \etal (2001). Since \gcs\ is purely
flux-limited, we avoid any bias that could be introduced in a subjectively
compiled sample, \eg\ based on the most preferred targets selected for deep
observations. To maintain the completeness of the sample, out of 63 systems
in \gcs, for two we have to use estimated temperatures based on the $L-T$
relation by Markevitch (1998). We have checked, however, that excluding
these two systems from the fit does not change the derived parameters of the
$M-T$ relation.

% Exclusion of the two systems does not change
% the fit results.}

% therefore we quote the fit results
% for the $M-T$ relation for the complete \gcs\ with 63
% clusters. Nevertheless,

% Construction of
% the flux-limited sample is complicated by absence of the temperature
% measurements for some of the systems. Limiting such a sample only to systems
% with measured temperatures has a drawback of having too few members. As a
% compromise, out of the 63 systems, for two we had to use estimated
% temperature based on $L-T$ relation by Markevitch (1999), resulting in
% corresponding temperatures of 2.2 and 5.6 keV for these systems. We have
% checked, however, that excluding these systems from the fit does not change
% the derived parameters of the $M-T$ relation.

% 
% The flux-limited sample contains two
% clusters for which no measured temperature has been found in the literature,
% however. For these the observational $L-T$ relation by Markevitch \etal
% (1998) has been used.  
In addition, we will use also a larger sample (enlarged \gcs, 88 clusters),
which is not strictly flux-limited, but contains only clusters with measured
temperatures.

A linear regression is performed in log($T$)-log($M$) space. The method
allows for errors in both variables and intrinsic scatter (Akritas \&
Bershady 1996). We use the bisector method to determine the best-fit
parameters (Isobe \etal 1990).  Errors are transformed into log space by
$\Delta\log(x)=\log(e)\, (x^+-x^-)/(2\, x)$ where $x^+$ and $x^-$ denote the
upper and lower boundary of the quantity's error range, respectively. The
values of the fit parameters
% for fits of the form $\log(M)=A+B\,\log(T)$
are given in Table \ref{tab:dhn}.

\input{thomas_tabs4.tex}

The results, obtained for the flux-limited and the enlarged sample agree
within the uncertainty of the fit. A correction for the observed redshift
does not result in any change in the derived parameters. The derived slope
of the $M-T$ relation is steeper than the value of 1.5, expected from
self-similar scaling relations.  The normalization obtained in the
hydrodynamical/$N$-body simulations of Evrard \etal (1996) is higher than
found here. No break in the $M-T$ relation is visible over the whole range
of temperatures.

Note, however, that the determination of the mass value itself depends on
the temperature, therefore some care has to be taken in the interpretation
of the fit results. In the discussion, which follows below, we take this
effect into account.

\begin{figure*}
\vspace*{-0.8cm}

\includegraphics[width=14.5cm]{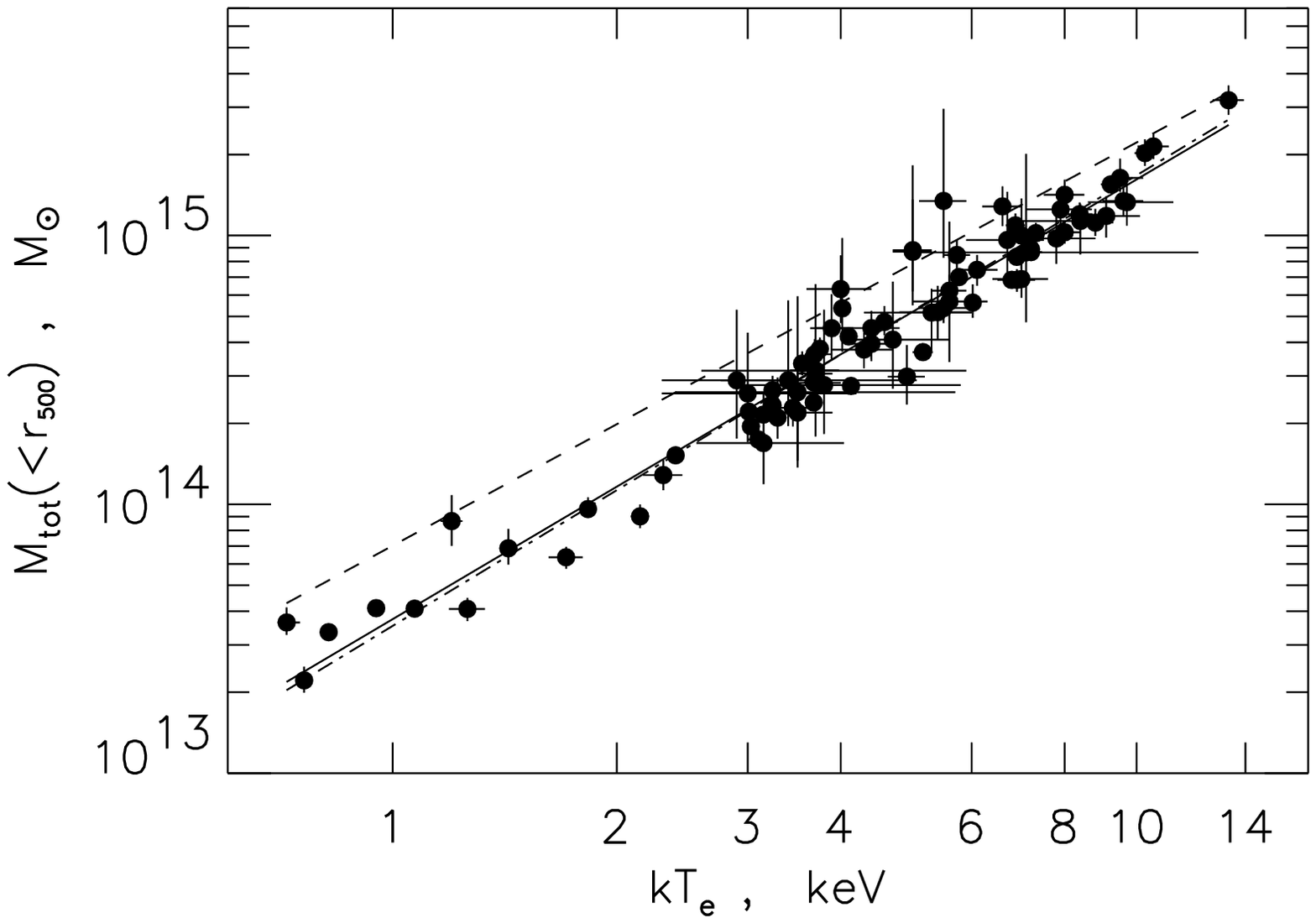}
\vspace*{-0.3cm}
\figcaption{$M-T$ relation for the enlarged \gcs\ (filled circles indicate
the data with solid line indicating the best-fit). For comparison, the fit
to the $M-T$ relation for the flux-limited \gcs\ is also shown (dot-dashed
line). The dashed line shows the result of simulations by Evrard \etal
(1996).
\label{fig:mtgcsz}
}

\includegraphics[width=14.8cm]{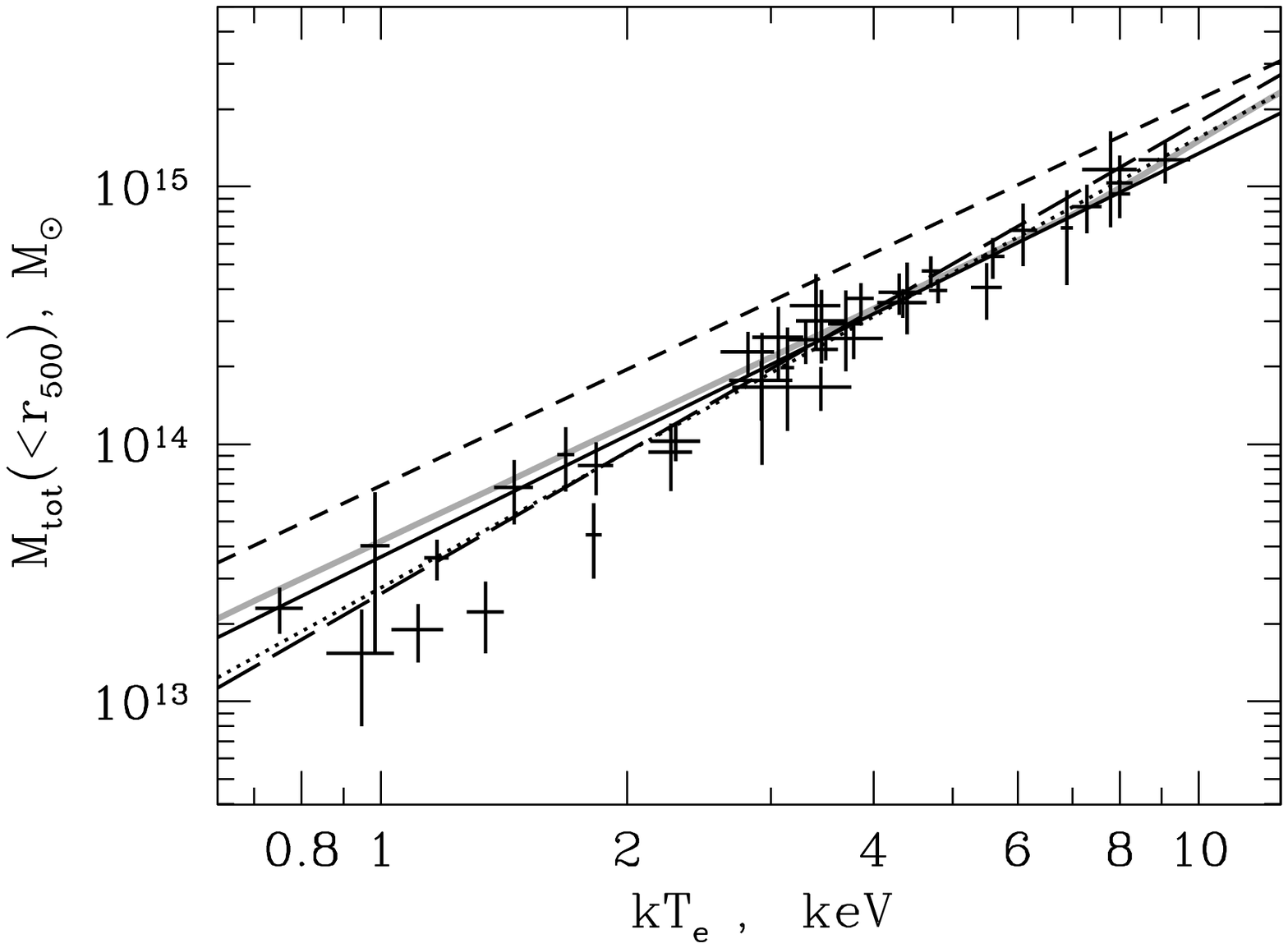} 
\vspace*{-0.7cm}
\figcaption{$M-T$ relation (analog to Fig.\ref{fig:mtgcsz}) for the sample
with temperature profiles. Crosses represent the mass determinations using
ASCA temperature profiles and ROSAT surface brightness profile fitting. The
dotted line denotes the best fit using the total sample, while the solid
line denotes the best fit, when groups ($M_{500}<5\times10^{13}$ \msun) are
excluded from the fitting. The dashed line shows the result of simulations
by Evrard \etal (1996). The long-dashed line shows the fit for the
low-temperature end of the sample ($kT_{ew}<4.5$ keV). The grey line shows
the effect of SN preheating on the $M-T$ relation, discussed in FAD.
\label{mt-clean}
}

\end{figure*}

\subsection{The sample with known temperature gradients.}

To derive the total masses for the clusters in this sample, we use the
spatially resolved temperature profiles found in ASCA measurements
(Markevitch \etal 1998, Finoguenov \& Ponman 1999, Finoguenov, David \&
Ponman 2000, hereafter FDP, FAD, Finoguenov, Jones \& B\"ohringer 2001). The
sample totals 39 systems with temperatures from 0.7 keV to 10 keV. This
corresponds to a factor of 100 difference in total mass, determined at a
given overdensity. A major difference of this sample, compared to studies of
Horner \etal (1999) and Nevalainen \etal (2000), is an inclusion of 22
systems with temperatures spanning the range from 0.7 keV to 3.5 keV. This
sample is therefore well suited to study the possible break in the $M-T$
relation, suggested by Nevalainen \etal (2000).

In calculating the total cluster masses we used polytropic indices to
describe the temperature profiles, omitting the cluster core, where effects
of cooling may be important.  For the total mass estimates, we used the fits
to the surface brightness profiles from ROSAT PSPC data on the outskirts of
the clusters from Vikhlinin \etal (1999), FDP, FAD, Finoguenov, Jones,
B\"ohringer (2001), thus avoiding the cooling zone of the cluster. We
estimate the uncertainty of surface brightness profile fitting on the mass
estimation as 4\% and propagate this error to the total mass. The
uncertainties in the total mass estimations are much larger and are due to
the uncertainty in temperature estimates and temperature gradients.

There could be a systematic effect in the analysis of ROSAT surface
brightness profiles of groups, caused by variation of the ROSAT
countrate-to-emission measure conversion factor at low emission temperatures
(B\"ohringer 1996; Finoguenov \etal 1999). To evaluate this effect, we
compare our fits to ROSAT surface brightness profiles with the spectral
normalizations, derived in our 3d modeling of ASCA data. The latter accounts
for both temperature and metallicity effects, but has larger systematics due
to the complex PSF. From a good agreement, found in this comparison, a
possible systematics in the total mass calculation, resulting from our
approach to measuring the $\beta$ values, is constrained to be less than
10\%.

Under the assumption of hydrostatic equilibrium, when the density profile is
described by a $\beta$-model and the temperature distribution is expressed
in polytropic form ($T(r)\propto n_{gas}^{\gamma-1}$), the total mass within
the radius $r=x r_c$ is simply

\begin{equation}\label{eq-mpc}
M_{tot}(<r) = 3.70\times10^{13} M_{\odot} T(r) r {3 \beta \gamma x^2 \over
  1 + x^2 }
\end{equation}

We employed Eq.\ref{eq-mpc} for our mass measurements. In Table \ref{tab:mass}
we list our mass determinations at overdensity 500.  Columns denote system
(1), redshift (2), emission weighted temperature (3), temperature at
$r_{500}$ (4), total mass within $r_{500}$ in $10^{14}$ \msun\ (5), measured
$r_{500}$ in Mpc (6), beta and core radius (7--8).  Temperature gradients
are tabulated in col. (9), expressed as polytropic indices ($\gamma$). In
col. (10) we cite the outer radii included in the analysis of ASCA data. All
errors in this table are given at the 68\% confidence level.
% {\bf We assume
%a 4\% error on total mass estimation resulting from measurement
%uncertainties in estimation of $\beta$ and core radii.}

In calculating the emission-averaged temperatures we removed the effects of
cooling flows and emission lines. Thus, these temperatures are not subject
to effects discussed in Mathiesen \& Evrard (2000, hereafter ME00). The
deviation of the measured $M-T$ relation relative to the simulated one could
be characterized by observing a {\it higher} temperature for a given
mass. If a distinction exists between the spectral temperature measurements
and the mass-averaged temperatures, as discussed in ME00, the
above-mentioned discrepancy should only increase. This, however, is not true
in the case of decreasing temperature profiles. We will return to this issue
below.

\subsection{$M-T$ relation from spatially resolved temperatures.}

In this section we investigate how the resulting parameters describing the
$M-T$ relation depend on the selection of the sample, with the most
important results listed in Table \ref{tab:dhn}.

Without the correction for the observed redshift the fit to the $M-T$
relation using the bisector method by Akritas \& Bershady (1996) gives

\[ M_{500}=(2.61_{-0.33}^{+0.38})10^{13} \times kT_{ew}^{1.78_{-0.09}^{+0.09}}. \]

All the errors for the $M-T$ relation reported in this section were
calculated from 10000 bootstrap realizations. The mass is expressed in
\msun\ and the temperature in keV.

\begin{figure}[ht]
\includegraphics[width=3.2in]{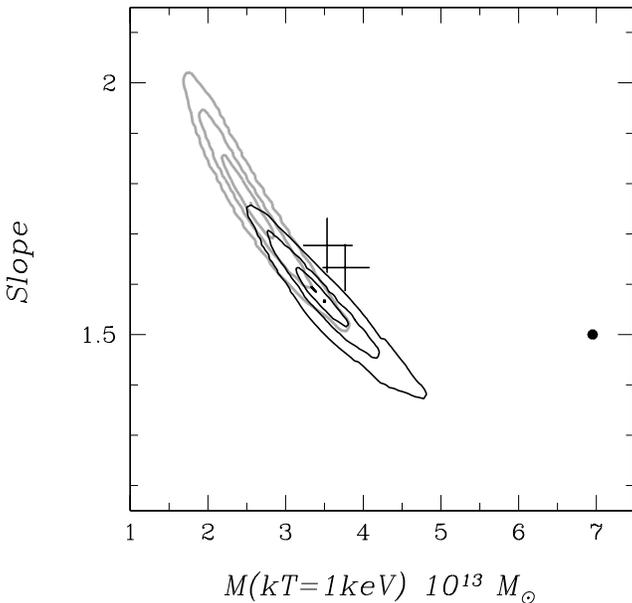}

\caption{Parameters of the bisector fit to the $M-T$ relation.  Contours,
drawn at 1,2 and 3 $\sigma$ confidence level, denote the parameters derived
from the sample with temperature profiles. In grey we show the fit to all
the data and in black those excluding the groups ($M_{500}<5\times10^{13}$
\msun). The black point shows the value obtained in simulations of Evrard
\etal (1996), while the black crosses denote the values obtained for the
enlarged and flux-limited \gcs\ sample with the error bars shown at the 68\%
confidence intervals.
\label{mt-parameters}
}
\end{figure}

Correcting for the redshift we obtain

\[ M_{500} =  (2.64_{-0.34}^{+0.39})10^{13} \times kT_{ew}^{1.78_{-0.09}^{+0.10}}. \]

Thus, there is no significant effect due to the redshift correction for our
sample, composed from nearby objects, and we will omit it in further
relations.

Our results on the slope of the $M-T$ relation are in agreement with
findings of Nevalainen \etal (2000), who determined a significantly steeper
slope than 1.5 ($1.79\pm0.14$, 90\% confidence interval is cited) at an
overdensity of 1000.

Horner \etal (1999) find a flatter $M-T$ relation, consistent with a value
of 1.5, but their sample lacks groups. Nevalainen \etal (2000) suggest on
the basis of their comparison of the $M-T$ relation, derived for hot
clusters and adding a few groups, that there might be a break in the $M-T$
relation, occurring below 4 keV. Since our data uniquely sample a
temperature range from groups to clusters of galaxies, we are in a position
to check this suggestion.

The $M-T$ relation derived without groups ($M_{500}>5\times10^{13}$ \msun) is
\[ M_{500} =  (3.57_{-0.35}^{+0.41})10^{13} \times kT_{ew}^{1.58_{-0.07}^{+0.06}} .\]

\includegraphics[width=3.2in]{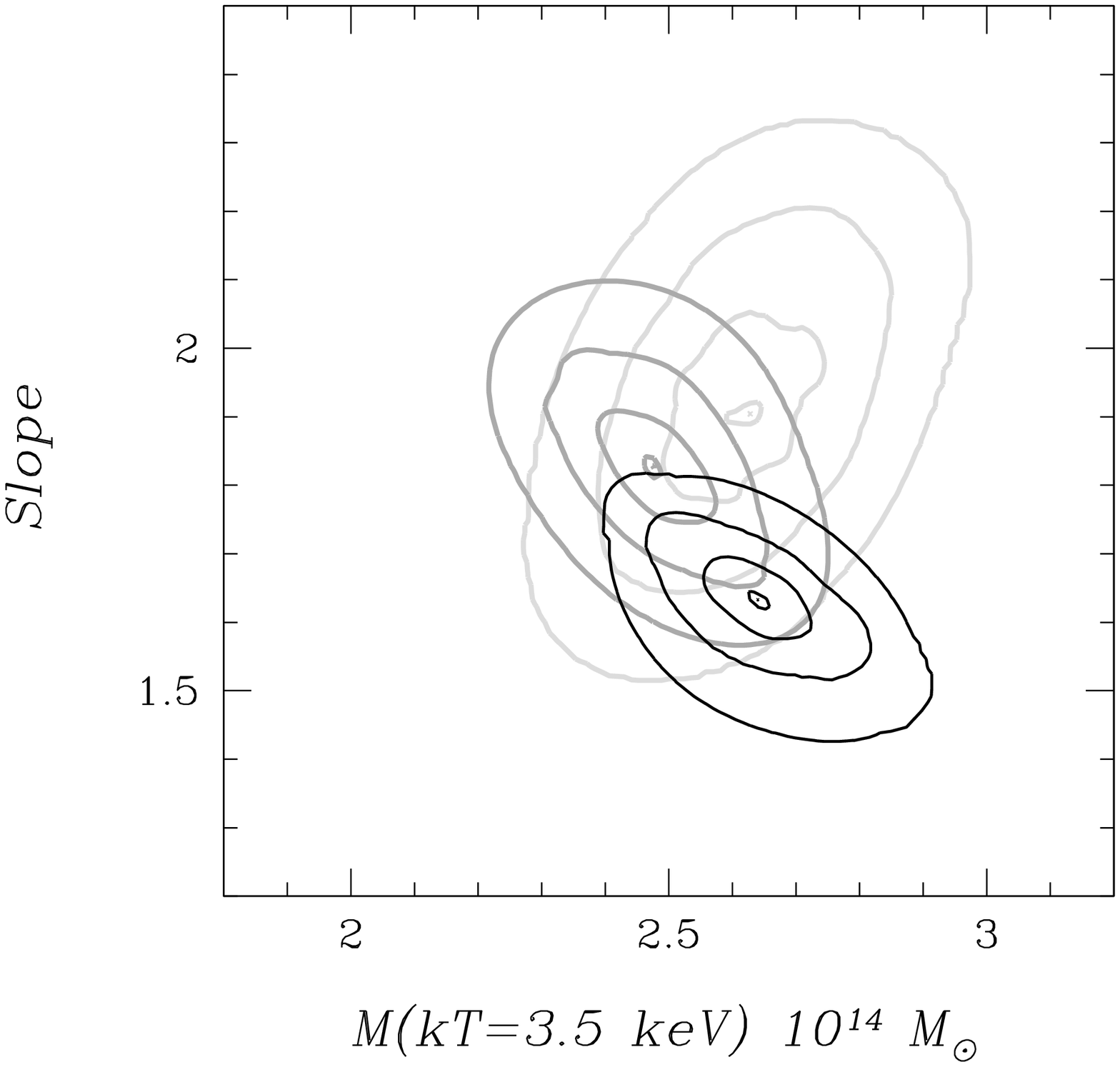}

\figcaption{Parameters of the bisector fit to the $M-T$ relation, derived
from the sample with temperature profiles. Light grey contours describe the
slope and normalization of the $M-T$ relation for clusters with temperature
below 4 keV. In grey we show the fit to all the data and in black those
excluding the groups ($M_{500}<5\times10^{13}$ \msun). Contours are drawn at
1,2 and 3 $\sigma$ confidence level. The innermost contour marks the center.
\label{mt-parameters2}
}

Note that it is more straightforward to operate in terms of mass in
separating the groups, since their temperatures are strongly affected by
even a slight degree of preheating (Loewenstein 2000; Tozzi, Scharf \&
Norman 2000).

Restricting ourselves to the systems with temperatures above 3 keV, the fit
is

\[ M_{500} =  (4.22_{-0.66}^{+0.85})10^{13} \times kT_{ew}^{1.48_{-0.12}^{+0.10}}.\]

The low-temperature end of the $M-T$ relation (systems with temperatures
below 4.5 keV) yields

\[ M_{500} =  (2.45_{-0.39}^{+0.44})10^{13} \times kT_{ew}^{1.87_{-0.14}^{+0.15}}. \]

The above fits and the data are shown in Fig.\ref{mt-clean}. The parameters
of the fits are compared with the results for the \gcs\ and simulations of
Evrard \etal (1996) in Fig.\ref{mt-parameters}.

To study the confidence area of the parameter estimation for the above fits,
we choose a normalization at 3.5 keV, which makes the determination of the
slope less dependent on the normalization. We present the values derived
this way in Fig.\ref{mt-parameters2}. It can be seen from this figure that
the total sample is inconsistent with a power law index of 1.5 on more than
99.9\% level. The high-temperature system sample (excluding groups),
although revealing a flatter index consistent with a value of 1.5, is not
strongly deviant from the total sample, \eg the break in the $M-T$ relation
has only 95\% confidence. A steeper slope, derived for the low-temperature
end of the sample, can still be considered as a fluctuation. However, larger
errors in the parameter determination in the case of inclusion of groups are
due to the large spread of groups on the $M-T$ relation (see
Fig.\ref{mt-clean}). So, the meaning of ``fluctuation'' is that a subsample
of systems leading to a derivation of the flat slope could be drawn from the
existing sample at a high probability. The origin of the scatter is further
discussed in Sec.\ref{sec:zf}.

In many studies the virial radius is suggested as a unit of length (\eg\
Evrard \etal 1996; Markevitch \etal 1998; Cen \& Ostriker 1999) to provide a
comparison among the systems at equal overdensity. For these estimations,
the luminosity averaged temperature of the cluster is used (Markevitch \etal
1998, FDP). Therefore, we provide here a relation between $r_{500}$ and the
luminosity-weighted X-ray temperature, derived from the data in Table
\ref{tab:mass}:
%   
% For practical purposes of computing the radius of overdensity from measured
% X-ray temperatures, we performed a fit to the observed $r_{500}$ dependence
% on $T$

\[ r_{500} =  0.63_{-0.01}^{+0.01} \times \sqrt{kT_{ew}}, \]

where $r_{500}$ is in Mpc and $kT_{ew}$ in keV. For a given temperature this
relation implies a 20\% smaller value for $r_{500}$ compared to similar
formulae derived from the simulations of Evrard \etal (1996).

\section{Polytropic vs Isothermal}\label{sec:poly}

As given in Eq.\ref{eq-mpc}, the deduced mass depends on the temperature and
the parameters $\beta$ and $\gamma$, describing the shape of the density and
temperature profile. To further elucidate the origin of the behavior of the
$M-T$ relation and to circumvent the direct dependence of $M$ on $T$ we
examine in Fig.\ref{fig-beta} the dependence of $\beta$ (correctly $\beta
x^2/(1+x^2)$ at $r_{500}$) on $T$. We note four systems with $\beta\sim0.3$
that imply a steep dependence of beta on the temperature. They are mainly
responsible for the steep slope of the $M-T$ relation in our sample with
spatially resolved temperatures. In fact, excluding these systems (IC4296,
HCG62, HCG51, NGC3258), but leaving other groups in, the $M-T$ relation is
given by
  
\[ M_{500} =  (3.57_{-0.26}^{+0.27})10^{13} \times kT_{ew}^{1.58_{-0.05}^{+0.05}}. \]

Excluding two more systems (A262, HCG94), whose $\beta$-index is less than
0.5, no further changes in the derived parameters of the $M-T$ relation are
obtained: 

\[ M_{500} =  (3.75_{-0.25}^{+0.29})10^{13} \times kT_{ew}^{1.56_{-0.05}^{+0.05}}. \]

All of the three groups in a sample of Nevalainen \etal (2000) have a
$\beta$-index less or equal to 0.5. The above demonstrates the importance of
selection effects on the derived slope of the $M-T$ relation, especially on
the group scale, originating from apparent scatter in the $M-T$ relation on
the scales of groups. Observationally, this scatter is correlated with flat
gas density profiles, which are often taken for signs of preheating (Metzler
\& Evrard 1997).  In reality, the situation is more complex, since as
pointed out by Loewenstein (2000), when the effect of the SN feedback is the
only one responsible for the observed $M-T$ relation, in low-mass systems, a
large amount of metals associated with SN explosions should be observed and
also isentropic cores, corresponding to the case of adiabatic
collapse. Although some of the groups do show such features (Finoguenov
\etal 2001), these are surprisingly not the most deviant systems in the
$M-T$ relation.

\includegraphics[width=3.2in]{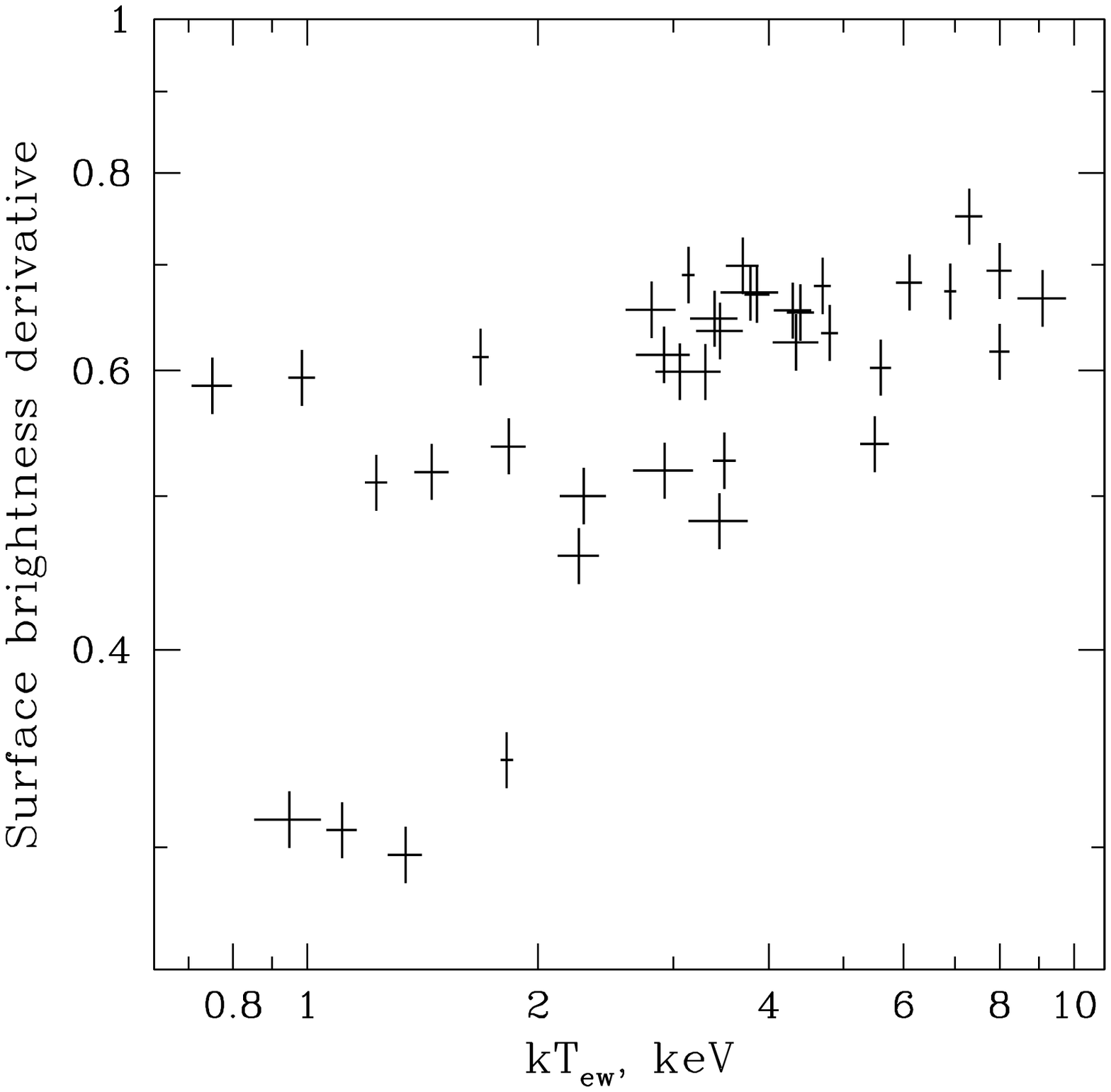}

\figcaption{Relation between density gradient (defined by $\beta
x^2/(1+x^2)$) and $T_{ew}$. 
\label{fig-beta}
}

Horner \etal (1999) have found that the dependence of $\beta$ on $T_{ew}$ is
responsible for the steepening of the derived $M-T$ in the isothermal
assumption, but when the temperature profiles are taken into account, the
slope becomes 3/2 again. To verify this, we rewrite Eq.\ref{eq-mpc} in terms
of overdensity

\begin{equation} \label{eq-dlt}
M(\delta, \beta, \gamma, T(\delta)) \approx 2.2 \times 10^{15} M_{\odot}
\delta^{-1/2} \beta^{3/2} \gamma^{3/2} T(\delta)^{3/2} ,
\end{equation}

where $\delta$ denotes the chosen overdensity. If the idea of Horner \etal
(1999) is correct, Eq.\ref{eq-dlt} should contain a term counterbalancing
the dependence of $\beta$ on $T_{ew}$, with possibilities $\gamma \propto
T_{ew}^{\alpha}$, $T(\delta)/T_{ew} \propto T_{ew}^{\alpha}$ or both. The
index $\alpha$ of this (counter-)dependence is $-0.26\pm0.03$ according to
Horner \etal (1999), but weaker dependences of $\beta$ on $T$ are also
reported (\eg\ $0.16$ in Vikhlinin \etal 1999).

As is seen from Figs.\ref{fig-gamma} and \ref{fig-terat}, the data show two
trends, which cancel each other: both very hot and very cold systems seem
to have stronger temperature gradients. So, the trend on the high-energy
part counterbalances the weak dependence of $\beta$ on T, while the trend
in the low-temperature part only reinforces the trend observed in
$\beta$. Thus, it becomes clear why inclusion of groups makes such a drastic
difference in the derived $M-T$ relation. An overall fit to these figures
gives $\gamma=(1.080\pm0.005)T_{ew}^{(-0.010\pm0.004)}$ and
$T(\delta)/T_{ew}=(0.88\pm0.18)T_{ew}^{(-0.03\pm0.14)}$.

\includegraphics[width=3.2in]{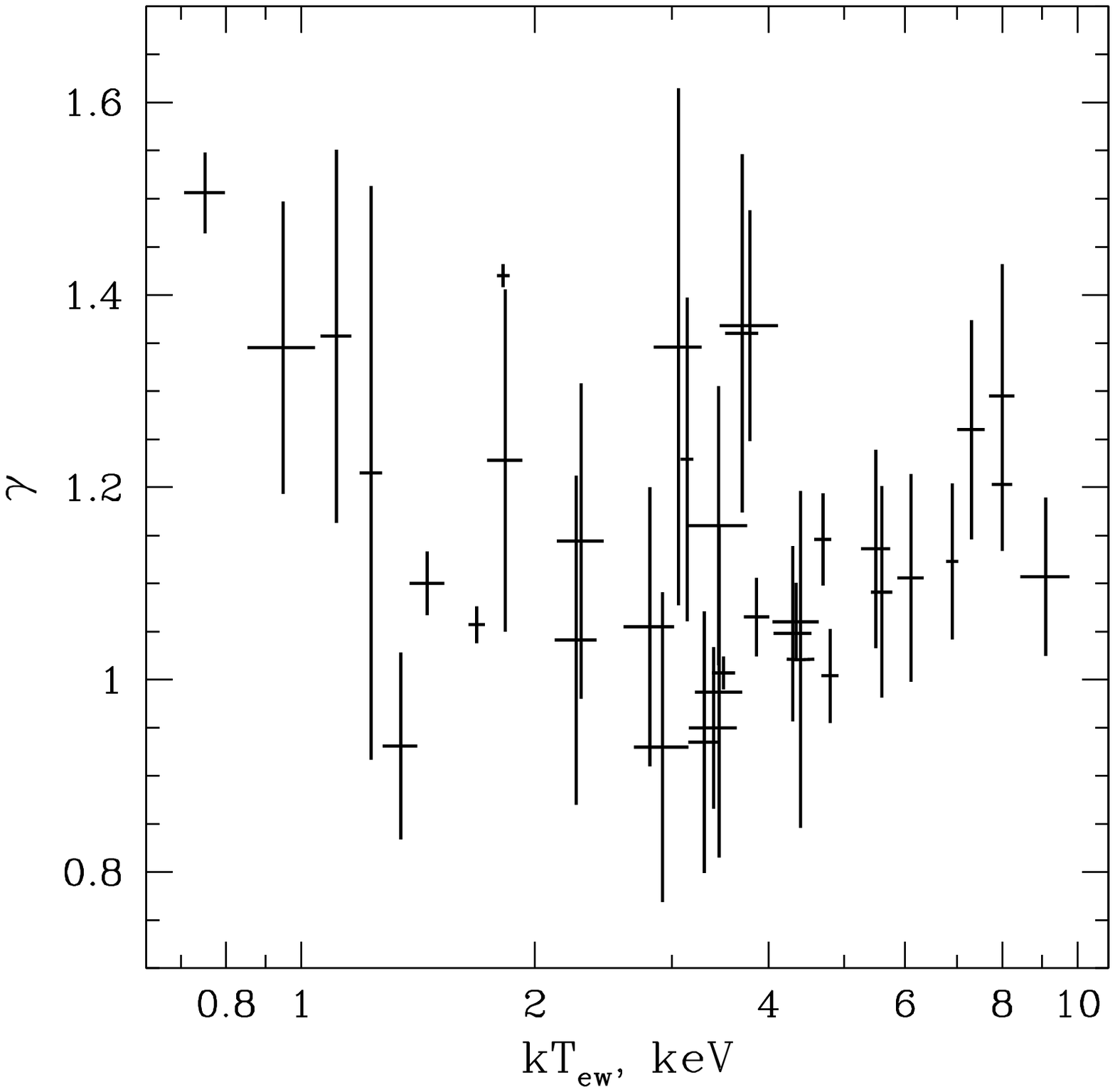}

\figcaption{Relation between $\gamma$ and $T_{ew}$. The data are shown as
crosses with error bars drawn at the 68\% confidence level.
\label{fig-gamma}
}

\includegraphics[width=3.2in]{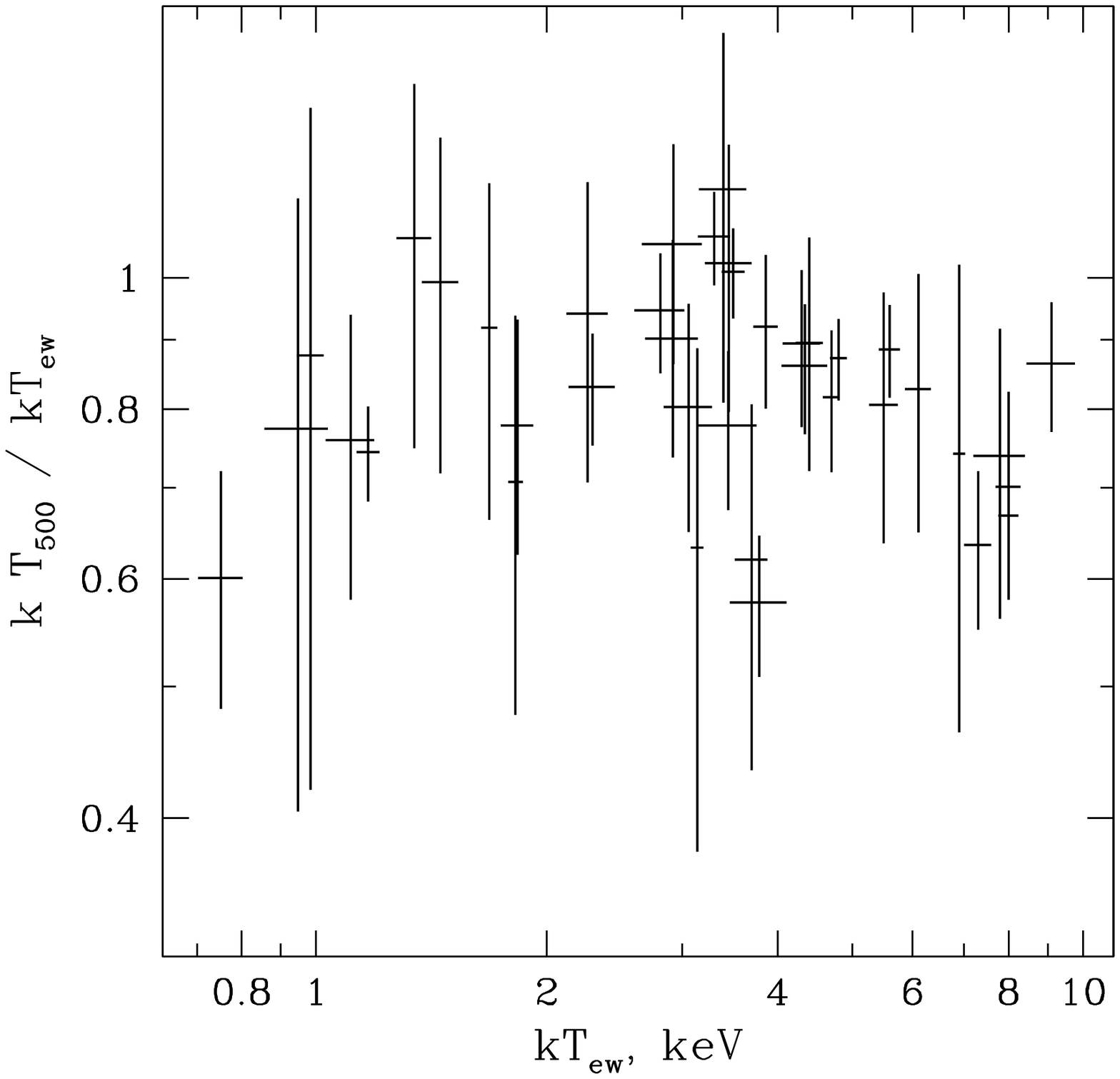}

\figcaption{Relation between $T_{500}/T_{ew}$  and $T_{ew}$. 
\label{fig-terat}
}

% 
% In Fig.\ref{fig-gamma} we study the first possibility. No significant trend
% is observed in the data with temperature profiles being marginally steeper
% for colder systems: $\gamma=(1.080\pm0.005)T_{ew}^{(-0.010\pm0.004)}$.
% 
% Also the $T(\delta)/T_{ew}$ does not show such a trend:
% $T(\delta)/T_{ew}=(0.88\pm0.18)T_{ew}^{(-0.03\pm0.14)}$.

% The data used for deriving these fits are shown in Figs.\ref{fig-gamma} and
% \ref{fig-terat}.  So it is not surprising that Nevalainen \etal (2000) did
% not find self-similar slope for the $M-T$ relation by combining the
% data on the hot clusters with a few groups. In fact, Figs.\ref{fig-gamma}
% and \ref{fig-terat} exhibit two trends, which average each other: both very
% hot and very cold systems seem to have stronger temperature gradients. So,
% the trend on the high-energy part, counterbalances the week dependence of
% $\beta$ on T, while the trend in the low-temperature part reinforces the
% picture. Thus it becomes clear why inclusion of groups makes such a drastic
% difference in the derived $M-T$ relation.

In fact, the arguments presented in Horner \etal (1999) are not strictly
correct. They base their conclusion on the fact that

$M(\beta)/M(True) \propto r^{0.4}$ for an individual cluster, where $M(\beta)$
is the mass estimate using the $\beta$-model and assuming isothermality and
$M(True)$ is the true mass of the cluster.

However, what is relevant is:

$M(\beta)/M(True) \propto (r/r_{core})^{0.4}$  for the cluster sample,

$r_{core} \propto r(\delta) \propto T^{0.5} $ where $r_{core}$ is the
cluster core radius and $r(\delta)$ is the radius of overdensity $\delta$,

from which it follows:

$M(\beta)/M(True) \propto (r(\delta)/r_{core})^{0.4} \propto
(T^{0.5}/T^{0.5})^{0.4} = constant.$

This means that if clusters are self similar, the overestimate of the beta
model is the same for all clusters. In view of this consideration, we note
that there is a very close agreement between the parameters of the $M-T$
relation determined in our two samples. This agreement, which demonstrates
that usage of the isothermality assumption does not bias the derived
parameters of the $M-T$ relation, can be used to justify the validity of the
$M-T$ relation, derived for high-redshift samples, where detailed
temperature measurements are difficult and an assumption of isothermality is
the only choice.

Following a suggestion of the referee, we examine also the effect of
assuming isothermality in deriving the total mass for the sample with
spatially resolved temperature measurements. The results are listed in Table
\ref{tab:dhn}. The slope of the $M-T$ relation, obtained for the total
sample, is still steeper, while avoiding the systems with flat $\beta$ gives
results consistent with \gcs\ in both slope and normalization. We note that
the lowest $\beta$ value for the \gcs\ sample is 0.44, which again supports
the idea of the importance of the selection of the systems with respect to
their values of $\beta$.

Concluding this section, we identify the inclusion of systems with low
values of $\beta$ as the most important cause of the steep slope of the
$M-T$ relation, which could be overcome by excluding such systems from the
sample. As we have shown, a flat $\beta$ is not necessarily a unique
characteristic of groups, which most likely implies a different importance
of preheating in low-mass systems. This is in qualitative agreement with the
preferential infall scenario, proposed by FAD.

In all the fits, however, the normalization of the $M-T$ relation appeares
smaller than in the simulations of Evrard \etal (1996). In the following, we
would like to comment on this issue. As pointed out by Nevalainen \etal
(2000), the resolution in the simulations of Evrard \etal (1996) is
insufficient to resolve the cluster cores. One can assume, however, that
their simulations are correct at low overdensities. Since we measure the
temperature up to the radius of the overdensity chosen for the mass
calculations, we can directly check this effect by using a temperature at
the radius of mass determination, ($T_{500}$), instead of the
luminosity-weighted temperatures. Such a comparison may also be less
affected by preheating, since at $\delta=500$ no significant variation in
the gas fraction is seen (Ettori \& Fabian 1999, Vikhlinin \etal
1999). Taking measured temperatures
% (see Table \ref{tab:mass} for a degree
% of extrapolation in determining $T_{500}$) 
also avoids many possible effects of averaging, discussed in ME00 and is
therefore closer to the relations predicted for the mass-weighted
temperature. A fit to the $M_{500}-T_{500}$ relation reads as

\[ M_{500} =  (3.29_{-0.59}^{+0.61})10^{13} \times kT_{500}^{1.89_{-0.14}^{+0.16}}, \]
(errors are stated at the 68\% confidence level).

The normalization of $M-T_{500}$ in the simulations by Evrard \etal (1996)
is about $9.-15.\times10^{13}$\msun\ (considering $T_{500}=0.6-0.8T_X$),
closer to the observed points, but still in obvious disagreement (fixing the
slope to 1.5 we obtain a normalization at 1 keV of
$4.81_{-0.29}^{+0.30}\times10^{13}$\msun).

\includegraphics[width=3.2in]{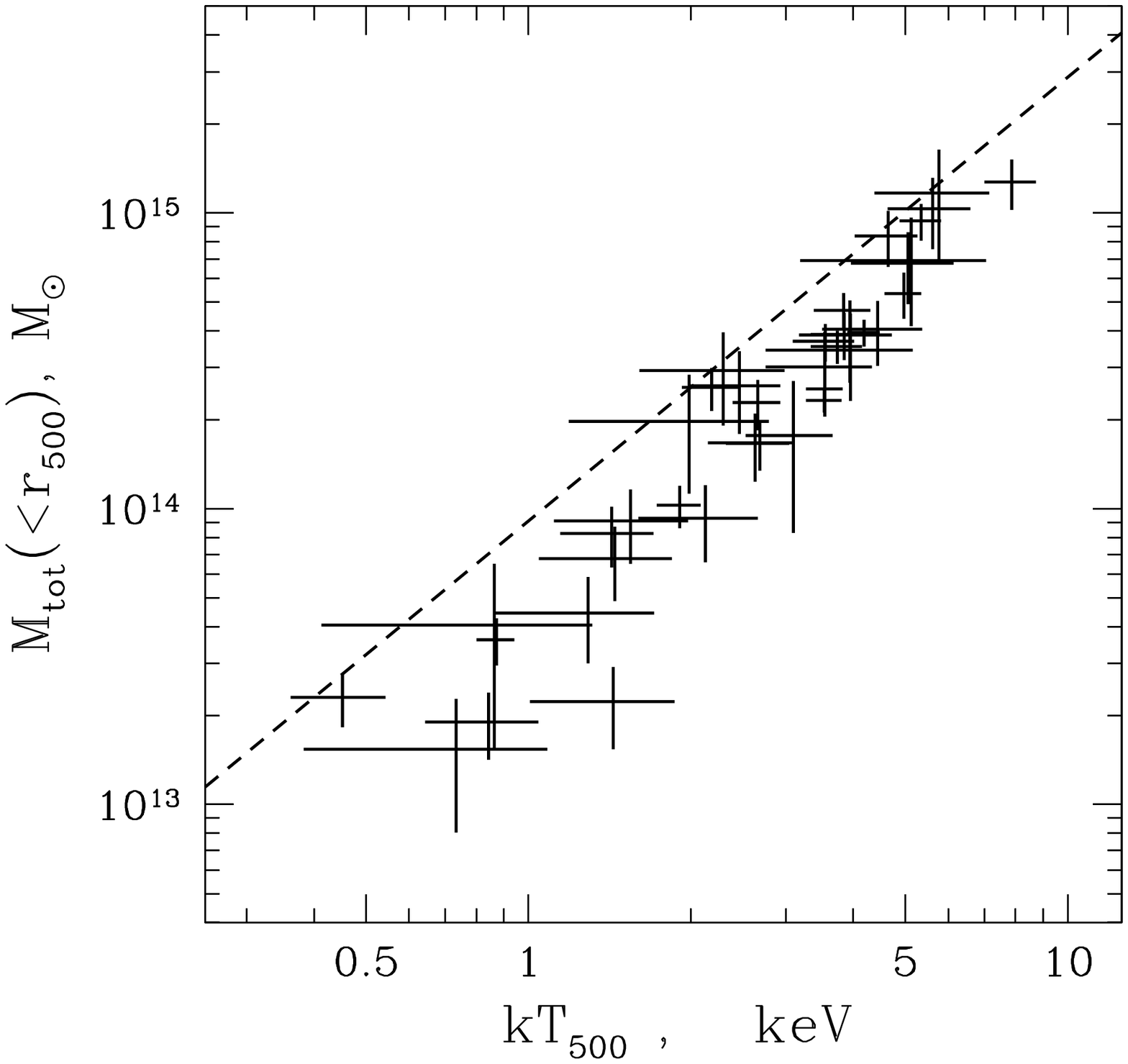}

\figcaption{$M_{500}-T_{500}$ relation. Dashed line shows the rescaled
  simulations by Evrard \etal (1996).
\label{mt-clean_t500}
}

% One can notice that hot clusters tend to come closer to the expected
% relation, while the cold systems remain aside. This is a result of effect of
% higher temperature gradients in hotter clusters. Since the temperature at
% this radius is more determined by enclosed mass, rather then subsequent
% adiabatic motion towards the center, the fact that some clusters satisfy the
% modeled relation means that their are formed very close to the moment of
% the observation. Since these are mostly the hottest systems, this statement
% is in agreement with general expectations of hierarchical clustering.
% Higher temperature gradients in hot cluster hence reflect a larger
% difference in the formation epoch between regions of equal overdensity,
% compared to cooler systems.

\section{Comparison with optical data}\label{sec:opt}

Comparison between the X-ray and the virial mass measurements, obtained
using velocity dispersions, is long known to be subject to contradictions,
with velocity bias considered as the most likely origin. Recently, a new
compilation of optical measurements was made by Girardi \etal (1998), where
a much more detailed study was carried out, \eg different velocity
dispersion profiles in clusters were identified. Since we have measured the
masses for many clusters in common, we can combine the X-ray mass with
velocity dispersion measurements to provide a comparison with
high-resolution dark matter simulations.

We take high-resolution $\Lambda CDM$ ($\Omega_{\rm M}=1-\Omega_{\Lambda}=0.3$,
$\sigma_8=1.0$; $H_{\rm o}=70$~km~s$^{-1}$~Mpc$^{-1}$) simulations using the
ART code (Kravtsov, Klypin, Khokhlov 1997), with $256^3$ particles of
$1.1\times10^9$ \msun\ each in a simulation box of $60h^{-1}$ Mpc (see
Gottloeber \etal 1999 for details). The particular runs of the ART code we
use for comparison are characterized by a scale of velocity dispersion with
overdensity of $\sigma_P \propto \delta^{0.06}$ with a residual scatter around
the best-fit of 20--30\%. Within the simulated box, 17 clusters have been
identified and we build the $M-\sigma_P$ relation, scaling the measurements
done at different overdensities. One possible weakness of such a comparison
is that in the simulations we have studied the dispersion of the dark
matter, not of the galaxies.

\includegraphics[width=3.2in]{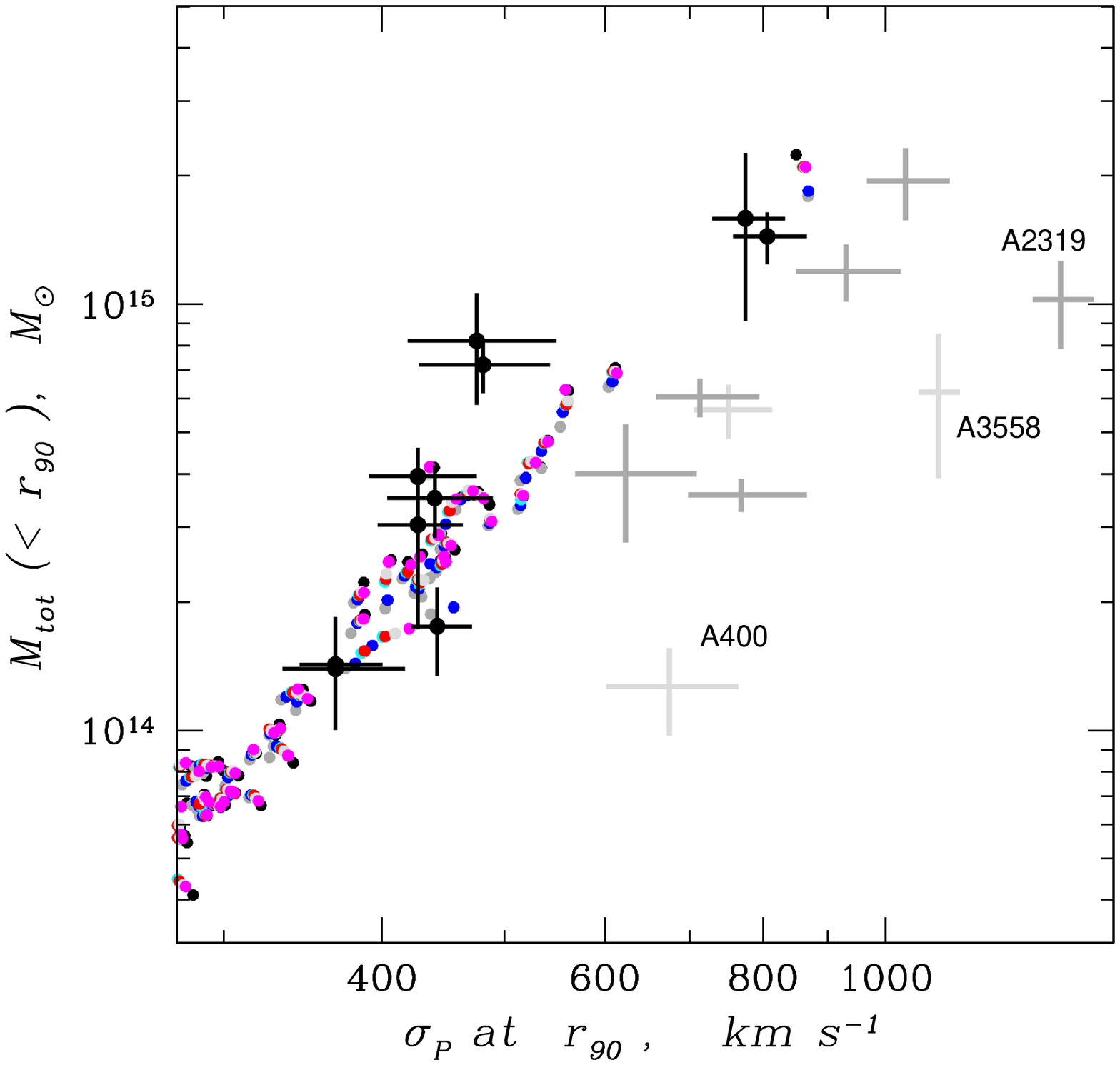}

\figcaption{$M-\sigma_P$ relation at overdensity of 90. Circles represent
the results of high-resolution simulations. Black crosses represent a
correlation of the X-ray mass measurements with the velocity dispersion of
galaxies {at overdensity of 90}. Type A clusters (Girardi \etal 1998) are
shown in black, type B in grey and C in light grey (this
identification has been made to denote different velocity dispersion
profiles measured in these clusters, which reflect, according to Girardi
\etal (1998), that type B and C cluster have been formed earlier and are
therefore more relaxed).
\label{m-s}
}

Optical observations reveal 3 types of clusters, according to their degree
of virialization. In comparing the data, we have scaled the optical velocity
dispersions according to the given type of cluster, using identification and
scaling profiles, reported in Girardi \etal (1998).  To compare our X-ray
results with the work of Girardi \etal (1998), we rescale our mass estimates
to the overdensity of 90, using the Navarro, Frenk \& White (1996) profile.

The resulting $M-\sigma_P$ relation is presented in Fig.\ref{m-s}. The
outliers on the $M-\sigma_P$ relation, A2319, A3558 and A400, are already
known to have a peculiar structure (Feretti, Giovannini, B\"ohringer 1997;
Venturi \etal 2000; Lloyd-Davies, Ponman, Cannon 2000). The scatter of the
other points around the simulations is within 30\% and therefore comparable
to the scatter seen in simulations. We note, however, that there is a trend
of type B and C clusters to have a higher velocity dispersion for a given
mass, which is consistent with a definition of their type: type B and C are
older systems, according to modeling of Girardi \etal (1998), that should
imply a higher formation redshift. Since such a correction is not introduced
into the virial mass calculation, this results in a slight bias in mass
estimates for these clusters, as seen in Fig.\ref{m-vir-xray}.

\includegraphics[width=3.2in]{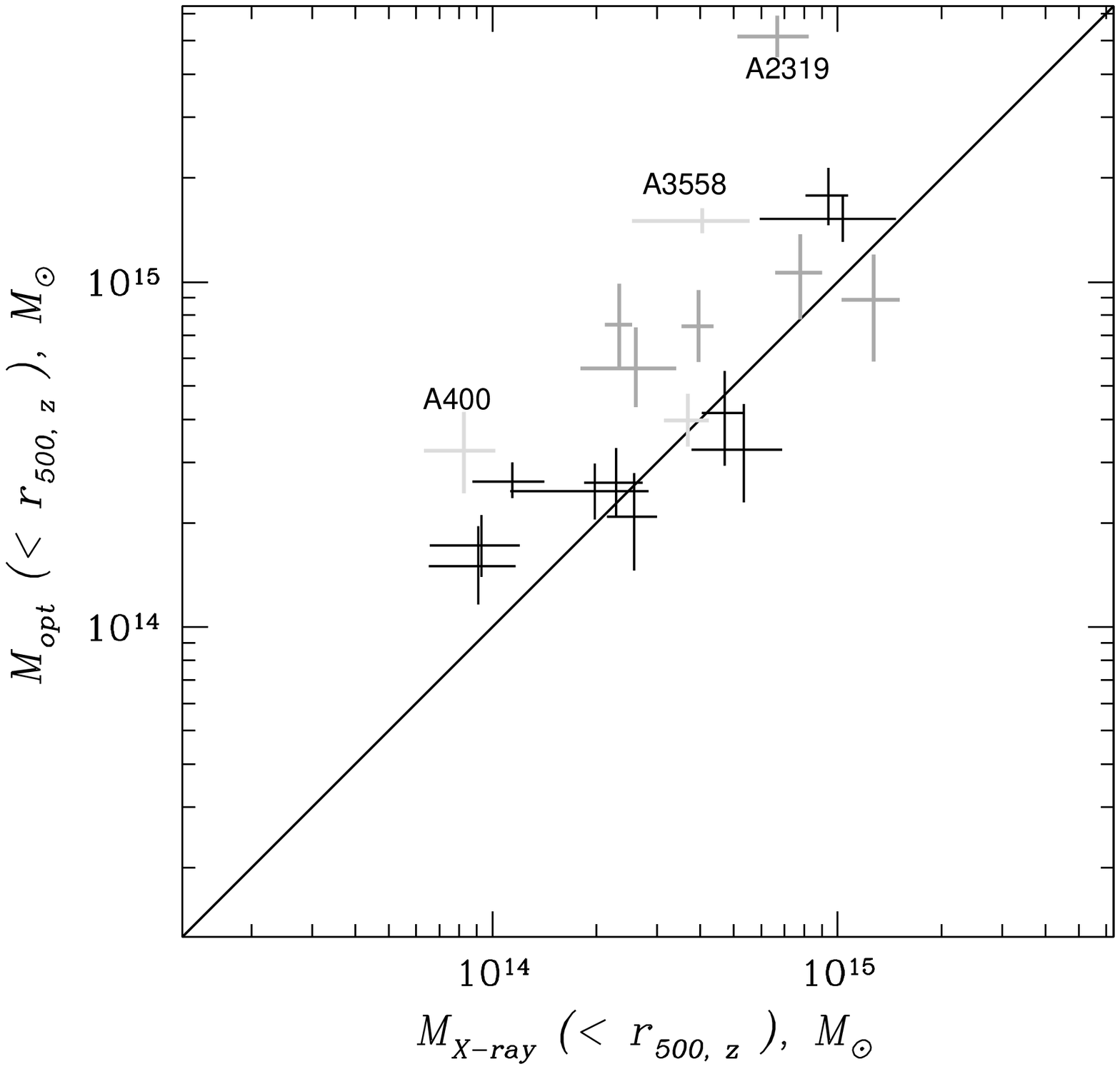}

\figcaption{Comparison of the X-ray and optical mass estimates. Black
crosses denote type A clusters in Girardi \etal (1998), grey crosses type B
and light grey crosses type C. Type A clusters show the best agreement with
the X-ray mass measurements.
\label{m-vir-xray}
}

\section{Implications on the Redshift (Epoch) of cluster formation}\label{sec:zf}

The issue of the redshift of cluster formation was strongly suggested in
studies of Lilje (1992), Kitayama \& Suto (1996), Voit \& Donahue
(1998). The effect on the $M-T$ relation is such that for a given mass, the
systems that formed at earlier times should have higher temperatures. Since
this scenario is in qualitative agreement with trends observed by comparing
our sample with the simulations of Evrard \etal (1996), we decided to
estimate whether the shift observed in the $M-T$ relation could be explained
just by this scenario.

To do this, we invert the problem, \ie\ use the X-ray mass and temperature
measurements and the $M-T$ relation obtained in the simulations to derive
the distribution of redshifts of cluster formation for our sample. In more
detail, for a measured mass of the system, we compare the measured
temperature with the value obtained from the simulation for the measured
mass and attribute the difference to the redshift of cluster formation,
which for $\Omega_{\rm M}=1$ is simply
$T_{observed}=T_{simulated}\times(1+z_f)$. To be able to do this, one should
make sure that the simulated relation explicitly assumes that the clusters
form at the redshift of observation. Fortunately, the simulations of Evrard
\etal (1996) have this assumption, which is quite logical for $\Omega_{\rm
M}=1$, used for most of their runs (Metzler, private communication).

In Fig.\ref{mt-z} we illustrate the effect of the redshift of cluster
formation, by comparison with the model of Lacey \& Cole (1993) following
the formulae presented in Balogh \etal (1999) for $\Omega_{\rm M}=0.3$. We
subdivide our sample into 3 parts with masses in the $0.1-0.8$, $0.8-3$ and
$3-15 \times10^{14}$ \msun\ range. From the Figure one can see that the
theoretical prediction varies significantly between the three subsets and
generally matches the trends seen in the corresponding subset. While a more
detailed comparison should await an $M-T$ relation simulated with much
better resolution, we can state already that a steeper slope of the observed
$M-T$ relation implies that lower-mass systems form preferentially at
earlier times, compared to rich clusters. The formation redshift
distribution for groups is wider than the model prediction, which can be
taken as a sign of the importance of SN preheating, but a more detailed
study is needed to verify this suggestion.

\includegraphics[width=3.2in]{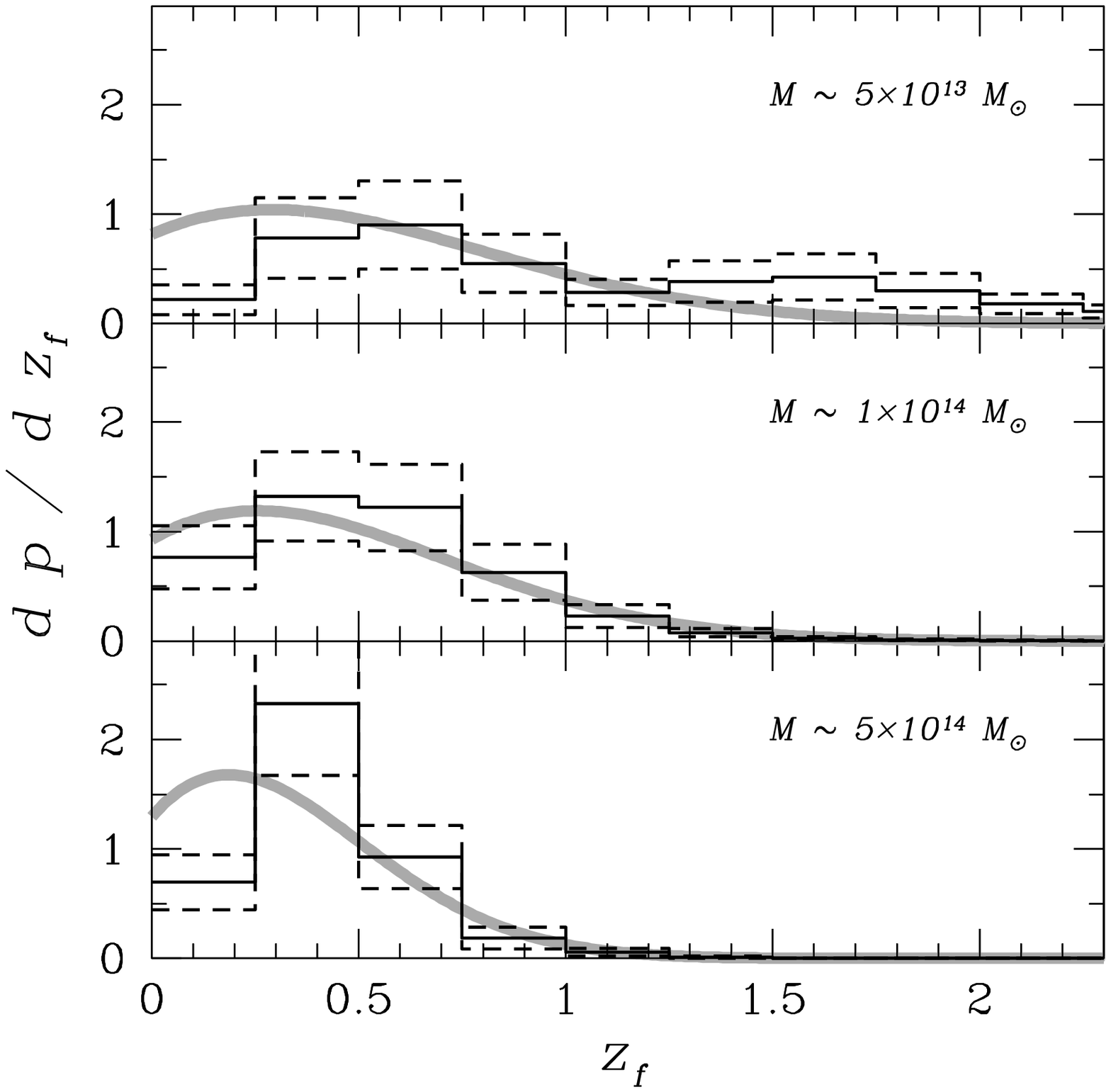}

\figcaption{Redshifts of cluster formation, deduced from a requirement for
measurements to match the simulations of Evrard \etal (1996). The black
solid histogram shows the calculation for the sample with temperature
profiles with dashed histograms indicating the number count errors at the 68\%
confidence level.  We present the comparison with the model of Lacey \& Cole
(1993) following the formulae presented in Balogh \etal (1999) for
$\Omega_{\rm M}=0.3$ and three mass ranges in our sample with typical masses
of $5\times10^{13}$ \msun, $1\times10^{14}$ \msun\ and $5\times10^{14}$
\msun\ (grey lines).
\label{mt-z}
}

We can already identify one source of bias in the formation redshift
distribution of the clusters in the present study. Systems that collapsed
recently bear indications of recent merger activity and therefore are
selectively removed from the sample aimed to determine the mass using the
assumption of hydrostatic equilibrium. Therefore, the first bin in the
derived $z_{f}$ distribution should be disregarded in
Fig.\ref{mt-z}. Flux-limited samples, such as \gcs, are in principle also
biased in this regard, since older systems are expected to be brighter for a
given mass.  The lack of clusters lying close to the simulations on the
$M-T$ plane is also caused by influence of a central region on the
determination of the luminosity averaged temperature, since formation of the
central part of a cluster is slightly shifted towards higher formation
epochs.
% Mathiesen (2000), suggests an effect of adiabatic increase of
% flux-weighted temperature after accretion of even a small mass fraction on
% the cluster. 

\section{Conclusions}\label{sec:sum}

We have studied the $M-T$ relation using a sample of clusters with resolved
temperature profiles and also using \gcs. Our findings are

\begin{itemize}

\item Using spatially resolved temperatures in mass estimates results in
similar parameters of the $M-T$ relation to the analysis assuming
isothermality. The observed slope of the $M-T$ relation is steeper than the
predictions of the self-similar relations and the measured normalization is
two times lower than that obtained in the simulations of Evrard \etal
(1996).

\item Significant scatter of the points on the $M-T$ relation was found for
groups. This is a likely source of the disagreement found between the
studies of Nevalainen \etal (2000) and Horner \etal (1999).

\item We find that the derived slope of the $M-T$ relation depends strongly
on the $\beta$ dependence on temperature and is not counterbalanced by
inclusion of temperature gradients, as proposed by Horner \etal (1999). This
provides an explanation for similar parameters of the $M-T$ relation,
derived using temperature profiles, compared to the method employing
the assumption of isothermality, as in the \gcs\ sample.

\item We conclude that the deviation of the measured $M-T$ relation from the
simulated one is due to the combined effect of preheating and the difference
between observed redshift and epoch of cluster formation. Avoiding clusters
with signs of recent merger activity in a sample selected for mass estimates
biases the sample toward earlier formation epochs.

\item To avoid the effects of preheating, we combine X-ray mass measurements
with galaxy velocity dispersion measurements of Girardi \etal (1998) to
build an $M-\sigma_P$ relation, which we compare with high-resolution dark
matter simulations. Apart from obvious outliers, also identified in other
studies, there is a clear difference between different cluster types,
derived according to velocity dispersion profile. In type B and C clusters,
identified as most relaxed, the velocity dispersion is highest for a given
mass, in qualitative agreement with predictions from models accounting for
the epoch of cluster formation.

\end{itemize}

\section*{Acknowledgments}

The authors thank Stephan Gottloeber for providing the results of
simulations using ART code. AF acknowledges useful conversations with
Monique Arnaud, Stefano Ettori, Yasushi Ikebe, Maxim Markevitch, Chris
Metzler, Volker Mueller and Alexey Vikhlinin during preparation of this
paper. AF acknowledges support from Alexander von Humboldt Stiftung and NASA
grant NAG5-3064 during preparation of this work. The authors thank the
anonymous referee for useful comments on the manuscript. The authors
acknowledge the devoted work of the ROSAT and ASCA operation and calibration
teams, without which this paper would not be possible.

\bibliographystyle{aabib99}
                                                           
% \bibliography{aamnem99,bibo_engl}
% \bibliographystyle{aabib99}

\clearpage

\input{tab_cl_clean2.tex}

\end{document}

%% file: mydefs.tex
\def\hea4{{\it HEAO~A4}}
\def\heaoa2{{\it HEAO~A2}}
\def\heao1{{\it HEAO~1}}

\def\eg{{\it e.g.}~}
\def\ie{{\it i.e.}~}

\def\h0{$H_{\rm o}=50$~km~s$^{-1}$~Mpc$^{-1}$}
\def\q0{$q_{\rm o}$}

\def\msun     {$M_{\odot}$}

\def\etal    {{ et~al.}~}

\def\cms3  {~{cm$^{-3}$}}

%% file: thomas_tabs4.tex
%-------Combined Thomas's Table-------
\begin{table}[ht]
{\renewcommand{\arraystretch}{1.3}
\caption{Fit parameter values for the $M-T$ relation$^{\dag}$.}\label{tab:dhn}
\begin{tabular}{lcc}
\hline
sample & slope & norm, 1 keV\\
& & $10^{13}$ \msun \\
%$B$ & $\Delta B$ & $A$ & $\Delta A$ \\
\hline
\multicolumn{3}{c}{\gcs} \\
\hline
flux-limited & $1.676\pm0.054$ & $3.53_{-0.30}^{+0.33}$\\
\hline
enlarged & $1.636\pm0.044$ & $3.74_{-0.27}^{+0.29}$\\
\hline
flux-limited, $\rhoc=\rhoc(z)$ & $1.679\pm0.054$ & $3.52_{-0.31}^{+0.33}$\\
\hline
enlarged, $\rhoc=\rhoc(z)$ & $1.636\pm0.046$ & $3.74_{-0.27}^{+0.29}$\\
\hline
\multicolumn{3}{c}{Sample with $kT$-profiles} \\
\hline
entire sample&$1.78_{-0.09}^{+0.09}$&$2.61_{-0.33}^{+0.38}$\\
\hline
entire sample, $\rhoc=\rhoc(z)$&$1.78_{-0.09}^{+0.10}$&$2.64_{-0.34}^{+0.39}$\\
\hline
$M>5\times10^{13}$\msun&$1.58_{-0.07}^{+0.06}$&$3.57_{-0.35}^{+0.41}$\\
\hline
$\beta>0.4$&$1.58_{-0.05}^{+0.05}$&$3.57_{-0.26}^{+0.27}$\\
\hline
\multicolumn{3}{c}{Implying isothermality to sample with $kT$-profiles} \\
\hline
entire sample&$1.89_{-0.09}^{+0.10}$&$2.45_{-0.32}^{+0.37}$\\
\hline
$M>5\times10^{13}$\msun&$1.74_{-0.06}^{+0.07}$&$3.04_{-0.29}^{+0.29}$\\
\hline
$\beta>0.4$&$1.66_{-0.04}^{+0.05}$&$3.50_{-0.23}^{+0.21}$\\
\hline
\end{tabular}
}
\begin{enumerate}
\item[{$^{\dag}$}]{\footnotesize ~ Errors are given at the 68\% confidence
level. A bootstrap method is used to estimate the errors.}
\end{enumerate}
\end{table}

%% file: tab_cl_clean2.tex
\begin{table*}
{
\begin{center}
\footnotesize
\caption{Mass determinations using spatially resolved temperatures.}\label{tab:mass}

\begin{tabular}{rcccrcccccc}
%\hline
\hline
Name   &   z      &    $kT$     & $kT_{500}$ & $M_{500}$ \ \ \ \ & $r_{500}$ & $\beta$ & $r_{core}$ & $\gamma$ &$R_{out}$ \\
       &          &     keV     & keV        & $10^{14}$ \msun    & Mpc       &      & Mpc       &  & Mpc      \\
\hline                                                                    
 A2029 & 0.077 & $9.10\pm0.66$ & $7.87\pm0.86$ & $12.71\pm2.44$ & $2.06\pm0.27$ & 0.68 & 0.28 & $1.11\pm0.08$ & 2.82 \\
A401   & 0.074 & $ 8.00\pm0.24$ & $ 5.34\pm0.47$ & $ 9.40\pm1.32$ & $ 1.70\pm0.16$ & 0.63 & 0.27 & $1.20\pm0.07$ & 1.82 \\
A3266  & 0.055 & $ 8.00\pm0.30$ & $ 5.61\pm0.98$ & $10.36\pm2.81$ & $ 1.93\pm0.36$ & 0.74 & 0.50 & $1.29\pm0.14$ & 1.57 \\
A1795  & 0.062 & $ 7.80\pm0.60$ & $ 5.77\pm1.39$ & $11.66\pm4.72$ & $ 2.00\pm0.56$ & 0.83 & 0.39 & $1.19\pm0.19$ & 1.56 \\
A2256  & 0.058 & $ 7.30\pm0.30$ & $ 4.64\pm0.62$ & $ 8.38\pm1.80$ & $ 1.79\pm0.27$ & 0.82 & 0.52 & $1.26\pm0.11$ & 1.83 \\
A3571  & 0.040 & $ 6.90\pm0.12$ & $ 5.12\pm1.93$ & $ 6.91\pm2.75$ & $ 1.68\pm0.46$ & 0.69 & 0.27 & $1.12\pm0.08$ & 2.28 \\
A1651  & 0.085 & $ 6.10\pm0.24$ & $ 5.05\pm1.09$ & $ 6.77\pm1.85$ & $ 1.67\pm0.32$ & 0.70 & 0.26 & $1.11\pm0.11$ & 1.62 \\
A119   & 0.044 & $ 5.60\pm0.18$ & $ 4.96\pm0.39$ & $ 5.34\pm0.95$ & $ 1.54\pm0.19$ & 0.66 & 0.48 & $1.09\pm0.11$ & 1.28 \\
A3558  & 0.048 & $ 5.50\pm0.24$ & $ 4.44\pm0.93$ & $ 4.05\pm1.01$ & $ 1.41\pm0.24$ & 0.55 & 0.19 & $1.14\pm0.10$ & 1.39 \\
 A2199 & 0.030 & $4.80\pm0.12$ & $4.19\pm0.29$ & $ 3.95\pm0.41$ & $1.40\pm0.10$ & 0.64 & 0.14 & $1.00\pm0.05$ & 1.25 \\
  A496 & 0.033 & $4.70\pm0.12$ & $3.84\pm0.46$ & $ 4.70\pm0.67$ & $1.48\pm0.15$ & 0.70 & 0.25 & $1.15\pm0.05$ & 1.32 \\
 A4059 & 0.048 & $4.40\pm0.18$ & $3.94\pm0.77$ & $ 3.87\pm1.20$ & $1.39\pm0.30$ & 0.67 & 0.22 & $1.02\pm0.17$ & 0.92 \\
 A3112 & 0.075 & $4.34\pm0.30$ & $3.74\pm0.41$ & $ 3.54\pm0.45$ & $1.35\pm0.12$ & 0.63 & 0.12 & $1.06\pm0.04$ & 1.74\\
 Hydra & 0.057 & $4.30\pm0.24$ & $3.85\pm0.51$ & $ 3.89\pm0.71$ & $1.39\pm0.18$ & 0.66 & 0.12 & $1.05\pm0.09$ & 1.35 \\
 A2063 & 0.035 & $3.86\pm0.14$ & $3.55\pm0.46$ & $ 3.68\pm0.54$ & $1.36\pm0.14$ & 0.69 & 0.22 & $1.07\pm0.04$ & 0.93 \\
 MKW3S & 0.045 & $3.79\pm0.32$ & $2.18\pm0.26$ & $ 2.57\pm0.43$ & $1.21\pm0.14$ & 0.71 & 0.30 & $1.37\pm0.12$ & 1.16 \\
 A2657 & 0.040 & $3.70\pm0.18$ & $2.30\pm0.69$ & $ 2.93\pm1.02$ & $1.27\pm0.31$ & 0.76 & 0.37 & $1.36\pm0.19$ & 1.17 \\
  AWM7 & 0.017 & $3.50\pm0.12$ & $3.54\pm0.27$ & $ 2.33\pm0.21$ & $1.17\pm0.07$ & 0.53 & 0.10 & $1.01\pm0.02$ & 0.75 \\
 A2052 & 0.035 & $3.46\pm0.24$ & $3.54\pm0.79$ & $ 3.02\pm0.96$ & $1.28\pm0.28$ & 0.64 & 0.10 & $0.99\pm0.17$ & 0.46 \\
 HCG94 & 0.042 & $3.45\pm0.31$ & $2.68\pm0.36$ & $ 1.67\pm0.32$ & $1.05\pm0.14$ & 0.48 & 0.08 & $1.16\pm0.14$ & 1.08 \\
2A0335 & 0.035 & $3.40\pm0.24$ & $3.95\pm1.20$ & $ 3.44\pm1.13$ & $1.34\pm0.30$ & 0.65 & 0.08 & $0.95\pm0.08$ & 0.92 \\
 A4038 & 0.028 & $3.31\pm0.16$ & $3.55\pm0.28$ & $ 2.55\pm0.50$ & $1.21\pm0.16$ & 0.61 & 0.16 & $0.94\pm0.14$ & 0.38 \\
 A1060 & 0.011 & $3.14\pm0.06$ & $1.99\pm0.80$ & $ 1.98\pm0.86$ & $1.11\pm0.33$ & 0.70 & 0.16 & $1.23\pm0.17$ & 0.31 \\
 A2634 & 0.031 & $3.06\pm0.22$ & $2.46\pm0.47$ & $ 2.60\pm0.81$ & $1.22\pm0.26$ & 0.69 & 0.45 & $1.35\pm0.27$ & 0.83 \\
  MKW9 & 0.040 & $2.92\pm0.26$ & $3.10\pm0.57$ & $ 1.77\pm0.94$ & $1.07\pm0.39$ & 0.52 & 0.05 & $0.97\pm0.45$ & 0.71 \\
  AWM4 & 0.032 & $2.92\pm0.23$ & $2.63\pm0.48$ & $ 1.67\pm0.43$ & $1.05\pm0.19$ & 0.62 & 0.11 & $0.93\pm0.16$ & 0.46 \\
  A539 & 0.029 & $2.81\pm0.21$ & $2.66\pm0.27$ & $ 2.28\pm0.44$ & $1.16\pm0.16$ & 0.69 & 0.25 & $1.05\pm0.14$ & 0.72 \\
 MKW4S & 0.028 & $2.29\pm0.16$ & $1.91\pm0.18$ & $ 1.03\pm0.17$ & $0.89\pm0.10$ & 0.51 & 0.12 & $1.14\pm0.16$ & 0.75 \\
  A262 & 0.016 & $2.26\pm0.14$ & $2.13\pm0.53$ & $ 0.93\pm0.27$ & $0.86\pm0.17$ & 0.46 & 0.06 & $1.04\pm0.17$ & 0.44 \\
  A400 & 0.024 & $1.83\pm0.09$ & $1.43\pm0.28$ & $ 0.82\pm0.19$ & $0.83\pm0.13$ & 0.56 & 0.18 & $1.23\pm0.18$ & 0.64 \\
 N3258 & 0.009 & $1.82\pm0.04$ & $1.29\pm0.42$ & $ 0.45\pm0.14$ & $0.67\pm0.15$ & 0.34 & 0.05 & $1.42\pm0.01$ & 0.57 \\
  MKW4 & 0.020 & $1.68\pm0.04$ & $1.55\pm0.43$ & $ 0.91\pm0.26$ & $0.86\pm0.17$ & 0.64 & 0.18 & $1.06\pm0.02$ & 0.66 \\
 N6329 & 0.028 & $1.45\pm0.08$ & $1.44\pm0.40$ & $ 0.68\pm0.19$ & $0.78\pm0.15$ & 0.53 & 0.12 & $1.10\pm0.03$ & 0.64\\
 HCG51 & 0.026 & $1.34\pm0.07$ & $1.44\pm0.43$ & $ 0.22\pm0.07$ & $0.54\pm0.11$ & 0.30 & 0.08 & $0.93\pm0.10$ & 0.69 \\
 N5044 & 0.009 & $1.23\pm0.04$ & $0.87\pm0.07$ & $ 0.36\pm0.07$ & $0.63\pm0.08$ & 0.51 & 0.01 & $1.22\pm0.30$ & 0.28 \\
 HCG62 & 0.014 & $1.11\pm0.05$ & $0.84\pm0.20$ & $ 0.19\pm0.05$ & $0.51\pm0.09$ & 0.31 & 0.02 & $1.36\pm0.19$ & 0.58 \\
 N4325 & 0.026 & $0.98\pm0.04$ & $0.86\pm0.45$ & $ 0.40\pm0.25$ & $0.65\pm0.28$ & 0.59 & 0.01 & $1.14\pm0.56$ & 0.60\\
IC4296 & 0.013 & $0.95\pm0.09$ & $0.73\pm0.35$ & $ 0.15\pm0.07$ & $0.47\pm0.16$ & 0.31 & 0.06 & $1.35\pm0.15$ & 0.32\\
 N5129 & 0.023 & $0.75\pm0.05$ & $0.45\pm0.09$ & $ 0.23\pm0.05$ & $0.54\pm0.08$ & 0.60 & 0.10 & $1.51\pm0.04$ & 0.58\\
%   hcg92  0.022  0.933 0.12 1.259 0.22 7.464E+13 1.630E+13   0.80   0.12  0.74 0.08 0.951  0.152  0.54
%   n2563  0.015  1.361 0.06 1.810 0.39 3.742E+13 8.258E+12   0.64   0.10  0.40 0.09 0.772  0.014  0.23
%    n507  0.016  1.337 0.03 1.966 0.35 5.393E+13 9.877E+12   0.72   0.09  0.44 0.02 0.815  0.009  0.22
%   n2300  0.006  1.008 0.14 2.260 1.12 2.540E+13 1.417E+13   0.56   0.22  0.33 0.03 0.572  0.262  0.22
%   n7619  0.013  0.997 0.01 1.224 0.38 1.568E+13 7.069E+12   0.48   0.15  0.33 0.05 0.776  0.523  0.21
\hline                      
\end{tabular}               
\end{center}                
                     
}                    
\end{table*}